\documentclass[letterpaper]{article}

\usepackage{emulateapj}
\usepackage{onecolfloat}
\usepackage{apjfonts}
\usepackage{graphicx}

\newcommand{\citealp}{\cite}

\slugcomment{Received 2000 February 9; accepted 2000 April 18}

\shortauthors{Fossati et al. }
\shorttitle{Evolution of Mkn~421: I. Timing Analysis}

\begin{document}
\twocolumn[
\title{X--ray Emission of Mkn~421: New Clues From Its Spectral Evolution. \\
I. Temporal Analysis }

\author{G. Fossati\altaffilmark{1,2}, 
A. Celotti\altaffilmark{3},
M. Chiaberge\altaffilmark{3},
Y. H. Zhang\altaffilmark{4},
L. Chiappetti\altaffilmark{4}, \\
G. Ghisellini\altaffilmark{5,6},
L. Maraschi\altaffilmark{5},
F. Tavecchio\altaffilmark{5},
E. Pian\altaffilmark{7},
A. Treves\altaffilmark{8} 
}

\begin{abstract}

Mkn~421 was repeatedly observed with \textit{Beppo}SAX in 1997--1998.
This is the first of two papers where we present the results of a thorough
temporal and spectral analysis of all the data available to us, focusing in
particular on the flare of April 1998, which was simultaneously observed
also at TeV energies.
Here we focus on the time analysis, while the spectral
analysis and physical interpretation are presented in the companion paper.
The detailed study of the flare in different energy bands reveals very
important new results: i) hard photons lag the soft ones by 2--3 ks --
a behavior opposite to what is normally found in high energy peak BL
Lacs X--ray spectra; ii) the flare light curve is symmetric in the
softest X--ray band, while it becomes increasingly asymmetric at
higher energies, with the decay being progressively slower than the
rise; iii) the flux decay of the flare can be intrinsically achromatic
if a stationary underlying emission component is present.
The temporal and spectral information obtained challenge the simplest
models currently adopted for the (synchrotron) emission and most
importantly provide clues on the particle acceleration process.  
\end{abstract}

\keywords{%
galaxies: active ---
BL Lacertae objects: general --- 
BL Lacertae objects: individual (Mkn 421) --- 
X--rays: galaxies ---
X--rays: general}
]
\altaffiltext{1}{Center for Astrophysics and Space Sciences, University of California at San Diego, 9500 Gilman Drive, La
Jolla, CA 92093-0424, U.S.A.}
\altaffiltext{2}{e--mail: gfossati@ucsd.edu}
\altaffiltext{3}{International School for Advanced Studies, via Beirut 2--4, 34014 Trieste, Italy}
\altaffiltext{4}{Istituto di Fisica Cosmica G.~Occhialini, via Bassini 15, 20133 Milano, Italy}
\altaffiltext{5}{Osservatorio Astronomico di Brera, via Brera 28, 20121 Milano, Italy}
\altaffiltext{6}{Osservatorio Astronomico di Brera, via Bianchi 46, 22055 Merate (Lecco), Italy}
\altaffiltext{7}{ITeSRE/CNR, via Gobetti 101, 40129 Bologna, Italy}
\altaffiltext{8}{Universit\`a dell'Insubria, via Lucini 3, 22100 Como, Italy}


\section{Introduction}
\label{sec:introduction}

Blazars are radio--loud AGNs characterized by strong variability,
large and variable polarization, and high luminosity.  Radio spectra
smoothly join the infrared-optical-UV ones.  These properties are
successfully interpreted in terms of synchrotron radiation produced in
relativistic jets and beamed into our direction due to plasma moving
relativistically close to the line of sight (e.g. Urry \& Padovani
\citealp{up95}).  Many blazars are also strong and variable sources of
GeV $\gamma$--rays, and in a few objects the spectrum extends up to
TeV energies.  The hard X-- to $\gamma$--ray radiation forms a separate
spectral component, with the luminosity peak located in the MeV--TeV
range.

The emission up to X--rays is thought to be due to synchrotron
radiation from high energy electrons in the jet, while it is likely
that $\gamma$-rays derive from the same electrons via inverse Compton
(IC) scattering of soft (IR--UV) photons --synchrotron or ambient soft
photons (e.g. Sikora, Begelman \& Rees~\citealp{sbr94}, Ghisellini et
al. \citealp{gg_sed98}).

The contributions of these two mechanisms characterize the average
blazar spectral energy distribution (SED), which typically shows two broad
peaks in a $\nu F_\nu$ representation (e.g. von Montigny et al.
\citealp{vmon95}; Sambruna, Maraschi \& Urry \citealp{smu96}; Fossati et
al. \citealp{fg_sed98}): the energies at which the peaks occur and their
relative intensity provide a powerful diagnostic tool to investigate
the properties of the emitting plasma, such as electron energies and
magnetic field (e.g. Ghisellini et al. \citealp{gg_sed98}).  
Moreover variability studies, both of single band and of simultaneous
multifrequencies data, constitute the most effective means to constrain
the emission mechanisms at work in these sources as well as the
geometry and modality of the energy
dissipation. The quality and amount of X--ray data on the brightest
sources start to allow us to perform a thorough temporal analysis 
as function of
energy and determine the spectral evolution with good temporal resolution.

In X--ray bright BL Lacs (HBL, from High-energy-peak-BL Lacs, Padovani
\& Giommi \citealp{pg95}) the synchrotron maximum (usually) occurs in the
soft-X--ray band, and the inverse Compton emission extends in some
cases to the TeV band where -- thanks to ground based Cherenkov
telescopes -- four sources have been detected up to now: Mkn~421
(Punch et al. \citealp{punch92}), Mkn~501 (Quinn et al. \citealp{quinn96}),
1ES~2344+514 (Catanese et al.  \citealp{catanese_2344_98}), and
PKS~2155--304 (Chadwick et al. \citealp{chadwick98}). 
If the interpretation of the SED properties in terms of synchrotron and IC
radiation is correct, a correlation between the X--ray and TeV emission is
expected.

\smallskip Mkn~421 ($z$ = 0.031) is the brightest BL Lac object at
X--ray and UV wavelengths and the first extragalactic source
discovered at TeV energies, where dramatic variability has been
observed with doubling times as short as 15 minutes (Gaidos et
al. \citealp{gaidos96}).  As such it was repeatedly observed with X--ray
satellites, including \textit{Beppo}SAX.  
Remarkable X--ray variability correlated with strong activity at TeV
energies has been found on different occasions (Macomb et
al.~\citealp{macomb95,macomb96}, Takahashi et al.~\citealp{takahashi96},
Fossati et al.~\citealp{fg_lincei98}, Maraschi et
al.~\citealp{maraschi_letter}).
In particular, the 1998 \textit{Beppo}SAX data presented here were
simultaneous with a large TeV flare detected by the Whipple Observatory 
(Maraschi et al.~\citealp{maraschi_letter}).

This paper is the first of two, which present 
the results of a uniform, detailed spectral and temporal analysis of
\textit{Beppo}SAX observations of Mkn~421 performed during 1997 and 1998. 
Here we focus on the data reduction and the timing analysis, and also 
discuss the results on the spectral variability derived from the different
properties of the flux variations in different energy bands.

The paper is organized as follows.  We briefly summarize
the characteristics of \textit{Beppo}SAX (\S\ref{sec:bepposax_overview}),
and introduce the observations studied (\S\ref{sec:observations}). 
We then address the temporal analysis of the variability, 
considering several energy bands and
comparing the light curve features by means of a few simple estimators
for the 1997 and 1998 observations
(\S\ref{sec:light_curves}).  The remarkable flare observed in 1998 is
the object of a further deeper analysis, reported in Section
\S\ref{sec:timing98}, focused on timescales and time lags.
Section~\ref{sec:disc:variability} contains a summary of the results of the
temporal analysis, preparing the ground for the comprehensive discussion
presented in Paper~II (Fossati et al. \citealp{fossati_II}). There they are 
considered 
together with the results of
the spectral analysis and thus used to constrain a scenario able to interpret
the complex spectral and temporal findings.

\section{\textit{Beppo}SAX overview}
\label{sec:bepposax_overview}

For an exhaustive description of the Italian/Dutch \textit{Beppo}SAX
mission we refer to Boella et al. (\citealp{boella97}) and references
therein.  The narrow field coaligned instrumentation (NFI) on
\textit{Beppo}SAX consists of a Low Energy Concentrator Spectrometer
(LECS), three Medium Energy Concentrator Spectrometers (MECS), a High
Pressure Gas Scintillation Proportional Counter (HPGSPC), and a
Phoswich Detector System (PDS).  The LECS and MECS have imaging
capabilities in the 0.1--10~keV and 1.3--10~keV energy band,
respectively, with energy resolution of 8\% at 6~keV.  At the same
energy, the angular resolution is about 1.2 arcmin (Half Power
Radius).  In the overlapping energy range the MECS effective area (150
cm$^2$) is $\sim$ 3 times that of the LECS.  Furthermore the exposure
time for the LECS is limited by stronger operational constraints to
avoid UV light contamination through the entrance window (LECS
instrument is operated during Earth dark time only).  The HPGSPC
covers the range 4--120~keV, and the PDS the range 13--300~keV. 
In the overlapping interval the HPGSPC has a better energy resolution than
the PDS, but it is less sensitive of both PDS and MECS.
Therefore, HPGSPC data will not be discussed in this paper.

The present analysis is based on the {\scshape SAXDAS} linearized
event files for the LECS and the three MECS experiments, together with
appropriate background event files, as produced at the
\textit{Beppo}SAX Science Data Center ({\scshape rev~0.2, 1.1 and
2.0}).  The PDS data reduction was performed using the XAS software
(Chiappetti \& Dal Fiume \citealp{chiappetti_dalfiume}) according to the
procedure described in Chiappetti et al. (\citealp{chiappetti_2155}).

\section{Observations}
\label{sec:observations}

Mkn~421 has been observed by \textit{Beppo}SAX in the springs of 1997 and 1998.
The journal of observations is given in Table~\ref{tab:obs_log_97_98}.

\subsection{1997}
\label{sec:observations:1997}

The 1997 observation comprised several pointings spanning the interval
between April 29$^\mathrm{th}$ and May 7$^\mathrm{th}$.  MECS data are
not available for May 7$^\mathrm{th}$ because of the failure of the
detector unit 1 on May 6$^\mathrm{th}$.  In this paper we will not
consider the LECS data of this last day of the 1997 campaign, because
unfortunately LECS data alone do not provide useful spectral and
variability information.  The \textit{net} exposure time (excluding
the May 7$^\mathrm{th}$ data) was $\sim$ 52 and $\sim$ 117 ks for LECS
and MECS respectively, while the on--source time coverage computed as
the sum of the observations between each T$_\mathrm{start}$
and T$_\mathrm{stop}$ is of about 58 hr.

Results on the first half of the 1997 campaign (April 29$^\mathrm{th}$
to May 1$^\mathrm{st}$) have been presented by Guainazzi et al.
(\citealp{guainazzi_mkn421}), where the details and motivation of the 
observations are given.  We re--analyzed those data along with the
new ones, applying the same techniques, in order to obtain a
homogeneous set of results, necessary for a direct comparison.

\begin{deluxetable}{cccccccc}
\tablecolumns{8}
\tablewidth{0pc}
\tablecaption{Journal of Observations\label{tab:obs_log_97_98}}
\tablehead{
\multicolumn{2}{c}{Date (UTC)} && \multicolumn{3}{c}{Net Exposure Time (ks)} && \colhead{\textit{Beppo}SAX archive \#} \\
\cline{1-2} \cline{4-6} \\
\colhead{Start} & \colhead{Stop} && \colhead{LECS} & \colhead{MECS} & \colhead{PDS\tablenotemark{a}} 
}
\startdata
1997/04/29:04:02 & 1997/04/29:14:42 &&    11.6 &    21.8 & 18.4 && 50032001 \\
1997/04/30:03:19 & 1997/04/30:14:42 &&    11.4 &    24.1 & 23.0 && 50032002 \\
1997/05/01:03:17 & 1997/05/01:14:42 &&    11.2 &    23.8 & 22.8 && 50032003 \\
1997/05/02:04:10 & 1997/05/02:09:41 && \phn4.4 &    11.4 & 11.0 && 50016002 \\
1997/05/03:03:24 & 1997/05/03:09:41 && \phn4.3 &    11.7 & 11.2 && 50016003 \\
1997/05/04:03:25 & 1997/05/04:09:45 && \phn4.9 &    12.2 & 11.6 && 50016004 \\
1997/05/05:03:32 & 1997/05/05:09:45 && \phn5.0 &    11.9 & 11.4 && 50016005 \\
\textit{1997/05/07:04:47} & \textit{1997/05/07:10:27} && \textit{\phn6.0} & 
  \nodata & \textit{13.1} && \textit{50016006}\tablenotemark{b} \\[0.15cm]
1998/04/21:01:52 & 1998/04/22:03:13 &&    23.6 & 29.6 & 42.2 && 50686002 \\
1998/04/23:00:27 & 1998/04/24:06:37 &&    27.2 & 34.7 & 50.4 && 50686001 \\
\enddata
\tablenotetext{a}{\footnotesize 
Net exposure time over half of the total area, because of the
rocking mode in which the instrument is operated.}
\tablenotetext{b}{\footnotesize 
On May 7$^\mathrm{th}$ 1997 MECS~1 had a fatal failure.
Only LECS and PDS data are available for that day, and they are not
discussed in the paper. }
\end{deluxetable}


\subsection{1998}
\label{sec:observations:1998}

In 1998 \textit{Beppo}SAX observed Mkn~421 as part of a long monitoring
campaign involving \textit{Beppo}SAX, ASCA (Takahashi, Madejski, \& Kubo
\citealp{takahashi99_veritas}), \textit{Rossi}XTE (Madejski et al., in
preparation), and coverage from ground based TeV observatories
(Maraschi et al. \citealp{maraschi_letter}).  
The \textit{Beppo}SAX observation comprised two distinct long pointings,
started respectively on April 21$^\mathrm{st}$ and 23$^\mathrm{rd}$. 
The total \textit{net} exposure time for MECS telescopes has been of
64 ks and the LECS one adds up to 50 ks.  
Hereinafter we will refer to the two 1998 pointings simply as April
21$^\mathrm{st}$ and April 23$^\mathrm{rd}$, leaving out the year.

The actual on--source time was 55.5 hr.  Unfortunately celestial and
satellite constraints in late April 1998 were such that in each orbit
there are two intervals during which the reconstructed attitude is
undefined.  One occurs when the source is occulted by the dark Earth,
but the second longer one occurs during source visibility periods.
Thus there are about 19 min per orbit when the NFIs are pointing at
the source, but there is a gap in attitude reconstruction despite the
high Earth elevation angle.  During these intervals the source
position is drifting along the $\pm$X detector axis, from a position
in the center to about 12 arcmin off, passing the strongback window
support.  These intervals are excluded by the LECS/MECS event file
generation.  Using the XAS software it is possible to accumulate MECS
images also during such intervals, showing that the source is actually
drifting.  The drift causes a strong energy dependent modulation,
clearly visible in light curves accumulated with XAS, and which would
be very difficult to recover, as requires to build a response matrix
integrated from different position dependent matrices weighted
according to the time spent in each position.  We then did not try to
recover these data for the analysis.

In the case of PDS the source remains inside the collimator flat--top
during the drift, and therefore we used for the accumulation all
events taken when the Earth elevation angle was greater than 3 degrees
(inclusive of attitude gaps).

\section{Temporal Analysis}
\label{sec:light_curves}

Light curves have been accumulated for different energy bands.  The
input photon lists were extracted from the full dataset selecting
events in a circular region centered on the position of the point
source.  The extraction radii are 8 and 6 arcmin for LECS and MECS
respectively, chosen to be large enough to collect most of the photons
in the whole energy range\footnote{For bright and soft sources -- this
matters especially because the LECS point spread function (PSF) gets
rapidly broad below the Carbon edge, i.e. E $\lesssim$ 0.3 keV -- it
is recommended to select quite large values. }: the used values ensure
that $<$ 5~\% of the photons are missed, at all energies.

For MECS data we considered the merged photon list of all the
available MECS units, which were 3 for the 1997 observations and only 2
(namely MECS units {\scshape 2} and {\scshape 3}) in 1998.  

The expected background contribution is less than a fraction of a
percent (typically of the order of 0.2--0.5~\%) and therefore the
light curves have not been background subtracted.

In order to maximize the information on the spectral/temporal
behavior, the choice of energy bands shall be such that they are as
independent as possible. This should take into account the
instrumental efficiencies and, at a certain level, the details of the
spectral shape.  The main trade--off is the number of photons in each
band, that should be large enough to obtain statistically meaningful
results.  Taking into account the main features of the LECS and MECS
effective areas (e.g. the Carbon edge at 0.29~keV which provides with
independent energy bands below and above this energy) and the
steepness of the Mkn~421 spectrum (above a few keV the available
photons quickly become insufficient), we determined the following
bands: 0.1--0.5~keV and 0.5--2~keV for LECS data, 2--3
and 4--6~keV for MECS data.  Their respective ``barycentric''
energies\footnote{These have been computed \textit{directly from the count
spectra} and thus they are not model dependent. Moreover, despite of the
observed spectral variability, their values do not change more than a few eV.} 
are 0.26, 1.16, 2.35, 4.76~keV, respectively, thus providing a factor $\sim
20$ leverage, useful to test the energy dependence of the variability
characteristics.

\subsection{1997}
\label{sec:light_curves:1997}

\begin{figure}[t]
\centerline{\includegraphics[width=0.90\linewidth]{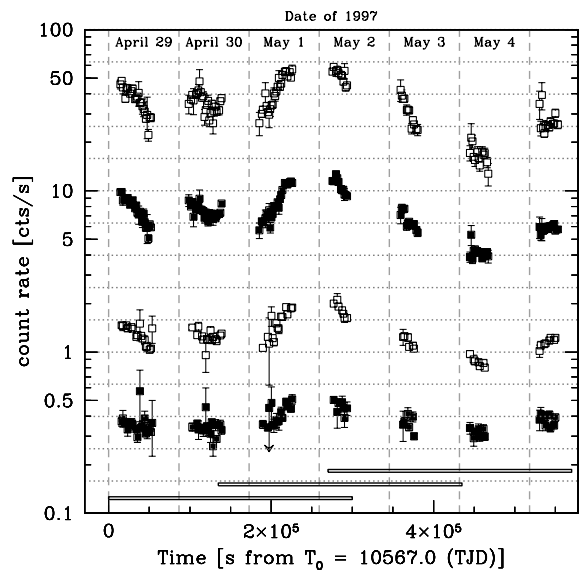}}
\vspace{-0.4cm}
\caption{\footnotesize\protect\baselineskip 8pt%
Light curves in four energy bands for the 1997 observations.  
They are shown in order of decreasing energy from top to bottom:
4--6~keV (multiplied by a factor 100),
2--3~keV (multiplied by a factor 10),
0.5--2~keV, 
0.1--0.5~keV.
The binning time is 1500~s.
The reference time TJD=10567.0 corresponds to 1997--April--29:00:00 UT.
The horizontal bars shown at the bottom of the plot mark the time intervals
considered for the analysis of the variability properties in
\S\ref{sec:light_curves:comparison_97_98}: subset--1 to --3 from left to right.
}
\label{fig:lc97}
\end{figure}

The four resulting light curves for 1997 are presented in
Figure~\ref{fig:lc97}. The source shows a high degree of flux
variability, with possibly a major flare between the third and fourth
pointings.   
The vertical scale is logarithmic allowing a direct
comparison of the amplitude of variations at different energies.
As anticipated we are not going to consider further the LECS
data of May 7$^\mathrm{th}$, however for sake of completeness we report
that the count rate was of 0.37$\pm$0.01 and 1.06$\pm$0.04 cts/s for the
0.1--0.5 and 0.5--2.0~keV energy bands respectively, comparable to the
level measured on May 5$^\mathrm{th}$.

The comparison of data with the overlaid reference grid shows that the
amplitude of variability is larger at higher energies, as commonly
observed in blazars at energies above the SED peak (e.g. Ulrich, Maraschi
\& Urry \citealp{umu97}).  
We will discuss this issue more quantitatively in section
\S\ref{sec:light_curves:comparison_97_98}.

\begin{figure}[t]
\centerline{\includegraphics[width=0.90\linewidth]{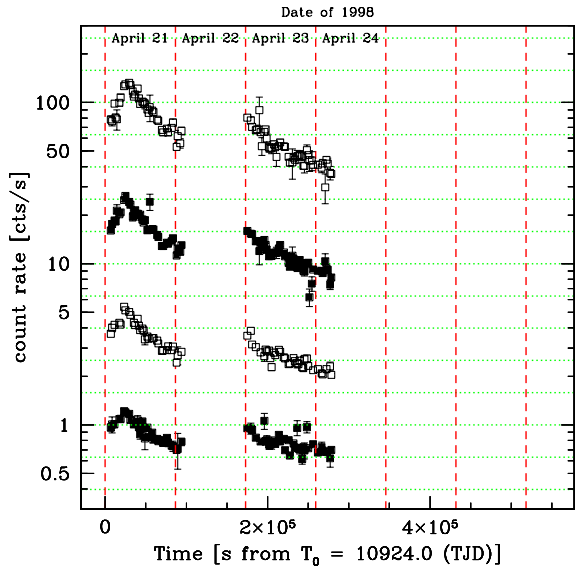}}
\vspace{-0.4cm}
\caption{\footnotesize\protect\baselineskip 8pt%
Light curves in four energy bands for the 1998 observations.  
They are shown in order of decreasing energy from top to bottom:
4--6~keV (multiplied by a factor 100),
2--3~keV (multiplied by a factor 10),
0.5--2~keV, 
0.1--0.5~keV.
The binning time is 1500 s.
The reference time TJD=10924.0 corresponds to 1998--April--21:00:00 UT.
In order to allow an easier comparison with the 1997 light
curves, the X scale has been expanded to the same length of that of
Figure~\ref{fig:lc97}, and also the Y--axis dynamic range is the same.  }
\label{fig:lc98}
\end{figure}

\begin{figure}[t]
\centerline{\includegraphics[width=0.95\linewidth]{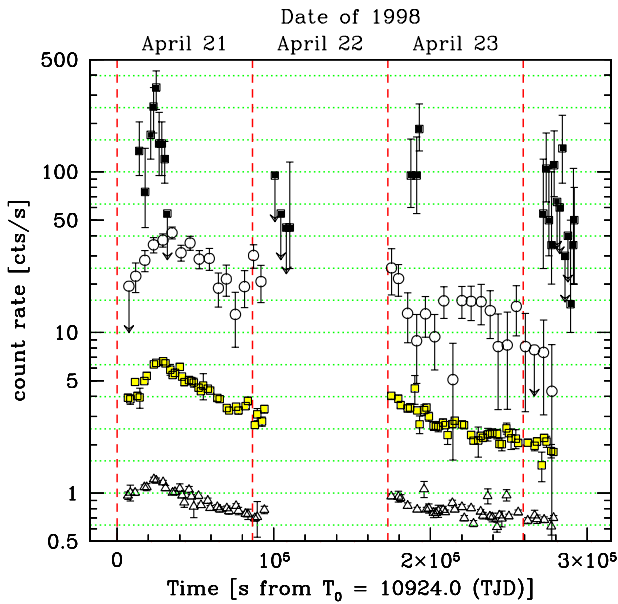}}
\vspace{-0.4cm}
\caption{\footnotesize\protect\baselineskip 8pt%
Light curves at TeV and X--ray energies, during the 1998 campaign.  
They are shown in order of decreasing energy from top to bottom:
Whipple E$\geq$ 2~TeV, (multiplied by a factor 500), 
PDS 12--26~keV (one point per orbit, multiplied by a factor 50), 
MECS 4--6~keV (multiplied by a factor 5) and LECS 0.1--0.5~keV (both
with 1500~s bins).
The count rate units are cts/s for \textit{Beppo}SAX data, and cts/min for
Whipple data. 
The thicker dashed grid marks the days, while the thinner ones have a 20 ks
spacing.
TJD=10924.0 corresponds to 1998--April--21:00:00 UT.
}
\label{fig:lc98tev}
\end{figure}

\subsection{1998}
\label{sec:light_curves:1998}
The 1998 light curves are shown in Figure~\ref{fig:lc98},
(see the caption for details about the axis scales). 
The overall appearance is somewhat different, being dominated by a single
isolated flare at the beginning of the campaign.  
The variability amplitude is similar to the 1997 one in each corresponding
energy band (see also \S\ref{sec:light_curves:comparison_97_98}).

The most striking and important result of the campaign
is that in correspondence with
the X--ray flare of April 21$^\mathrm{st}$ a sharp TeV flare was
detected by the Whipple Cherenkov Telescope, with amplitude of a
factor 4 and a halving time of about 2 hr (Maraschi et
al. \citealp{maraschi_letter}).  In Figure~\ref{fig:lc98tev} the Whipple
TeV (E$\geq$2~TeV) light curve is shown together with the
\textit{Beppo}SAX ones: LECS (0.1--0.5~keV), MECS (4--6~keV) and PDS
(12--26~keV) instruments. The peaks in the 0.1--0.5~keV, 4--6~keV and
2~TeV light curves are simultaneous \textit{within one hour}.  Note
however that the TeV variation appears to be both larger and faster
than the X--ray one. The 12--26~keV light curve shows a broader and
slightly delayed peak with respect to lower energy X--rays.  However
the very limited statistics does not allow us to better quantify this
event.  A more detailed account on this particular result is presented
and discussed by Maraschi et al. (\citealp{maraschi_letter}).

\begin{figure}[t]
\centerline{\includegraphics[width=0.90\linewidth]{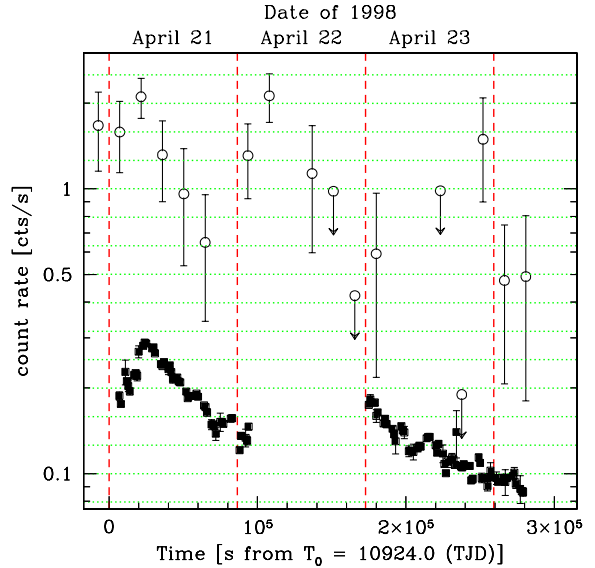}}
\vspace{-0.4cm}
\caption{\footnotesize\protect\baselineskip 8pt%
\textit{Rossi}XTE All Sky Monitor light curve (top) compared with the MECS
in the same (2--10~keV) band, scaled by a factor 1/20.}
\label{fig:lc98asm}
\end{figure}

It is also worth noticing that the positive (flux) offset of the
beginning of April 23$^\mathrm{rd}$ with respect to the end of the
April 21$^\mathrm{st}$ suggests the presence of a second flare
occurring between the two pointings.  
This seems to be also confirmed by the \textit{Rossi}XTE All Sky Monitor
(ASM) light curve, shown in Fig.~\ref{fig:lc98asm} together with the MECS
one in the (2--10~keV) nominal working energy range of the ASM.  
Single ASM $\sim$90 s dwells have been rebinned in 14400 s (i.e. 4 hr)
intervals, weighting the contribution of each single dwell on its
effective exposure time and on the quoted error. The MECS light curve
bin is 1000 s.  Indeed the ASM light curve appears to unveil the
presence of a second flare occurring in between the two
\textit{Beppo}SAX pointings, with a brightness level similar to that
of the detected one.

On the other hand the few Whipple data points at the time of
this putative second X--ray flare (at T = (100--120) $\times
10^3$ s, see Figure~\ref{fig:lc98tev}) do not indicate any (major) TeV
activity: the count rate measured by the Whipple telescope was in fact
significantly lower than that measured simultaneously with the first X--ray
flare.

\subsection{Comparison of 1997 and 1998 variability characteristics}
\label{sec:light_curves:comparison_97_98}

We can characterize the variability in different X--ray bands 
by two commonly used estimators: the fractional \textit{root mean
square} ($r.m.s.$) variability parameter F$_\mathrm{var}$, and the
\textit{minimum ``doubling/halving time''} T$_\mathrm{short}$ (for
definitions and details see Appendix~\ref{sec:appendix:estimators}, and
Zhang et al. \citealp{zhang_2155}).

\subsubsection{Fractional r.m.s. variability}

\noindent
For each (0.1--0.5, 0.5--2, 2--3 and 4--6 keV) band we
computed F$_\mathrm{var}$ for the whole 1997 and 1998 datasets which 
have the same on--source coverage (i.e. $\sim$ 55~hr).

Moreover, for the 1997 we considered three (partially overlapping)
subsets of data each spanning a time interval comparable to the length
of the whole 1998 observation, i.e. $\sim$ 300 ks.
The intervals cover the first 300~ks, a middle section, and the last
300~ks, as marked on Fig.~\ref{fig:lc97}.

\begin{figure}[t]
\centerline{\includegraphics[width=0.80\linewidth]{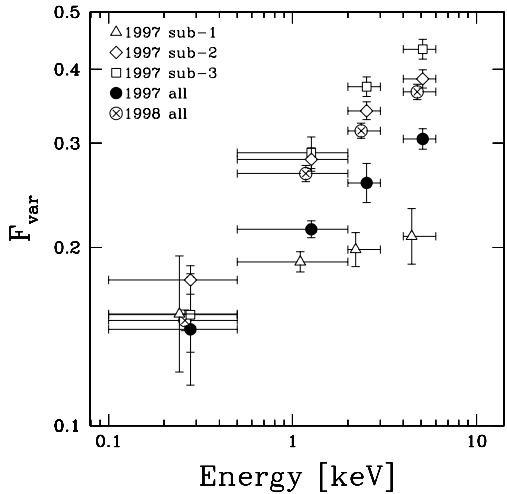}}
\vspace{-0.2cm}
\caption{\footnotesize\protect\baselineskip 8pt%
Plot of the \textit{root mean square} variability parameter,
F$_\mathrm{var}$, versus energy.
A small energy shift has been applied
for sake of clarity.  }
\label{fig:fvar_vs_energy}
\label{fig:fvarmod_vs_energy}
\end{figure}

For each dataset F$_\mathrm{var}$ is computed for 6 different light
curve binning intervals, namely 200, 500, 1000, 1500, 2000 and 2500 s,
and then the weighted average%
\footnote{To the F$_\mathrm{var}$ values obtained for each different time
binning, we assigned as weight $\equiv 1/\sigma^2$, where $\sigma$ is their
respective one sigma uncertainty.} 
of these values is computed.  
The results are listed in Table~\ref{tab:lc_fvar_t2} and shown in
Figure~\ref{fig:fvar_vs_energy}.

\begin{deluxetable}{lcccc}
\tablecolumns{5}
\tablewidth{0pc}
\tablecaption{Basic Variability Estimators\label{tab:lc_fvar_t2}}
\tablehead{
\colhead{}&
\colhead{LECS 0.1--0.5}&
\colhead{LECS 0.5--2}&
\colhead{MECS 2--3}&
\colhead{MECS 4--6}
}
\startdata
Energy (keV)& 0.26 & 1.18 &  2.36 & 4.76 \\
\cutinhead{\textit{r.m.s.} variability estimator F$_\mathrm{var}$} \\
1998 All    & $0.151 \pm 0.006$ & $0.267 \pm 0.009$ & $0.315 \pm 0.009$ & $0.366 \pm 0.011$ \\[0.1cm]
1997 All    & $0.145 \pm 0.035$ & $0.215 \pm 0.007$ & $0.257 \pm 0.020$ & $0.305 \pm 0.012$ \\[0.1cm]
1997 sub--1 & $0.154 \pm 0.039$ & $0.189 \pm 0.008$ & $0.198 \pm 0.014$ & $0.209 \pm 0.024$ \\[0.1cm]
1997 sub--2 & $0.176 \pm 0.010$ & $0.282 \pm 0.013$ & $0.340 \pm 0.012$ & $0.385 \pm 0.014$ \\[0.1cm]
1997 sub--3 & $0.154 \pm 0.024$ & $0.289 \pm 0.018$ & $0.374 \pm 0.015$ & $0.432 \pm 0.017$ \\[0.1cm]
  %
  %
  %
\cutinhead{\textit{shortest timescale} T$_\mathrm{short}$ (ks)} 
{1997: all data \hfill}           &  $80.6 \pm 46.6$     & $28.6 \pm \phn6.4$  & $25.5 \pm 2.4$ & $25.5 \pm 2.0$ \\[0.1cm]
{\phantom{1997:} halving \hfill}  & $167.4 \pm 31.0\phn$ & $62.9 \pm \phn4.4$  & $31.0 \pm 3.1$ & $32.6 \pm 8.0$ \\[0.1cm]
{\phantom{1997:} doubling \hfill} &  $81.7 \pm 62.1$     & $28.2 \pm \phn6.6$  & $27.3 \pm 4.1$ & $28.7 \pm 3.4$ \\[0.1cm]
{1998: all data \hfill}           &  $63.8 \pm 12.1$     & $28.1 \pm \phn4.2$  & $27.3 \pm 2.4$ & $21.4 \pm 1.5$ \\[0.1cm]
{\phantom{1998:} halving \hfill}  &  $63.8 \pm 12.1$     & $35.7 \pm \phn4.1$  & $36.9 \pm 8.7$ & $34.5 \pm 2.4$ \\[0.1cm]
{\phantom{1998:} doubling \hfill} & \nodata\tablenotemark{a} & $30.4 \pm 12.1$ & $28.4 \pm 3.3$ & $23.6 \pm 1.9$ \\
\enddata
\tablenotetext{a}{\footnotesize 
It has not been possible to derive a reliable estimate for this case due to insufficient data.}
\end{deluxetable}

If we treat all the datasets as independent and momentarily \textit{do not}
consider the \textit{subset--1} of the 1997 light curve (empty triangles in
Fig.~\ref{fig:fvar_vs_energy}) on which we will comment later, it appears
that:
\begin{enumerate}
\item In terms of \textit{root mean square} variability the lower energy
flux is less variable than the higher energy one, as already pointed out 
in section \S\ref{sec:light_curves:1997}.  This
holds independent of both the length of the time interval and the state
of the source.  
\item There is no relation between the brightness level and the
variability amplitude, since although in 1998 the source was at least
a factor 2 brighter than in 1997 the corresponding F$_\mathrm{var}$ are not
larger.
\end{enumerate}

What is different for the subset--1 of the 1997 light curve ? 

First two caveats on F$_\mathrm{var}$.  This quantity is basically
only sensitive to the average excursion around the mean flux, and does
not carry any information about either duty cycle or the actual
flux excursion.  Therefore in order for F$_\mathrm{var}$ to
meaningfully represent the variability, the ``extrema'' of the source
variability have to be sampled.  This is clearly not the case for the
first 1997 sub--light curve (see Figure~\ref{fig:lc97}).
Furthermore, F$_\mathrm{var}$ is estimated from the comparison of the
variances of the light curve and the measurements, but while the
latter ones probably obey a Gaussian distribution and are much smaller
than the former ones, the ``probability'' for the source to show a
significant deviation from the average (i.e. the count rate histogram
distribution) is in most cases highly non--Gaussian.  For a typical
``flaring'' light curve this has a ``core'' at low count rates,
resulting from the long(er) time spent by the source in between
flares, and a very extended tail to high(er) rates. The implicit
assumption of Gaussian variability can thus affect the estimate of
F$_\mathrm{var}$ (as for large amplitude and low duty cycle
variability).

F$_\mathrm{var}$ is thus a particularly poor indicator especially for
observations with a window function like the one of 1997, for which the
source temporal covering fraction is of the same order of the single
outburst, making likely to miss the flare peak, and the total span of
the observation does not significantly exceed the frequency of peaks.
We therefore believe this affects the results relative to the first subsample
of 1997.

\smallskip
In order to obtain a variability estimator more sensitive to the
dynamic range towards higher count rates (flaring) we also considered
a modified definition of it, F$_\mathrm{var,med}$: to study only the
width (amplitude) of the brighter tail only data
above the \textit{median} count rate are considered to compute
variance and expected variance. 
Note that the modified definition is still affected by the problem of
``representativity'' of all the source states.  
Since the median is always smaller than the average (for all energies, and
for all trial time binnings), the values of F$_\mathrm{var,med}$ 
are larger than F$_\mathrm{var}$, but in every other respect the results
are qualitatively unchanged.

\begin{figure}[t]
\centerline{\includegraphics[width=0.75\linewidth]{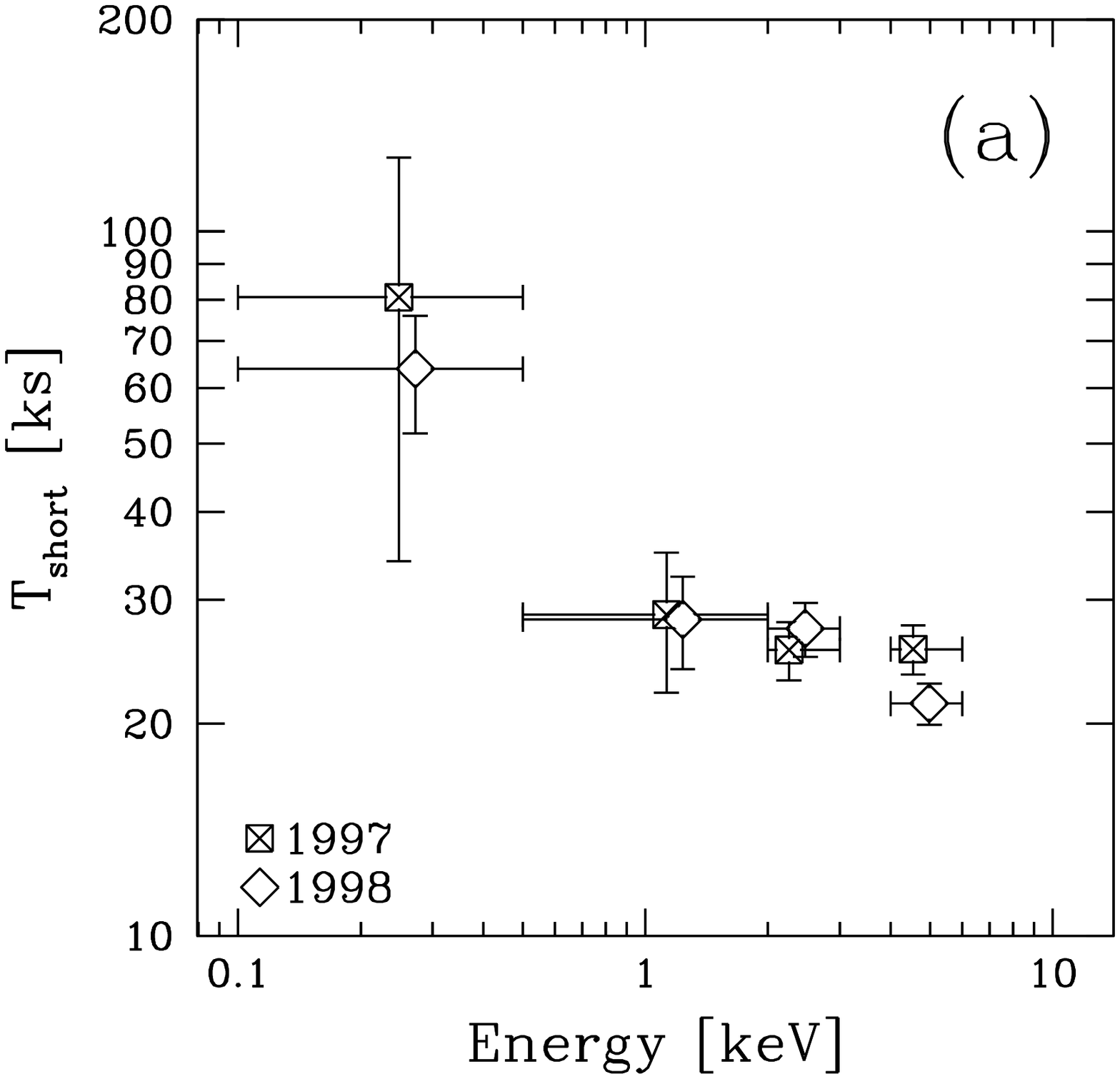}}
\vspace{-1.0cm}
\centerline{\includegraphics[width=0.75\linewidth]{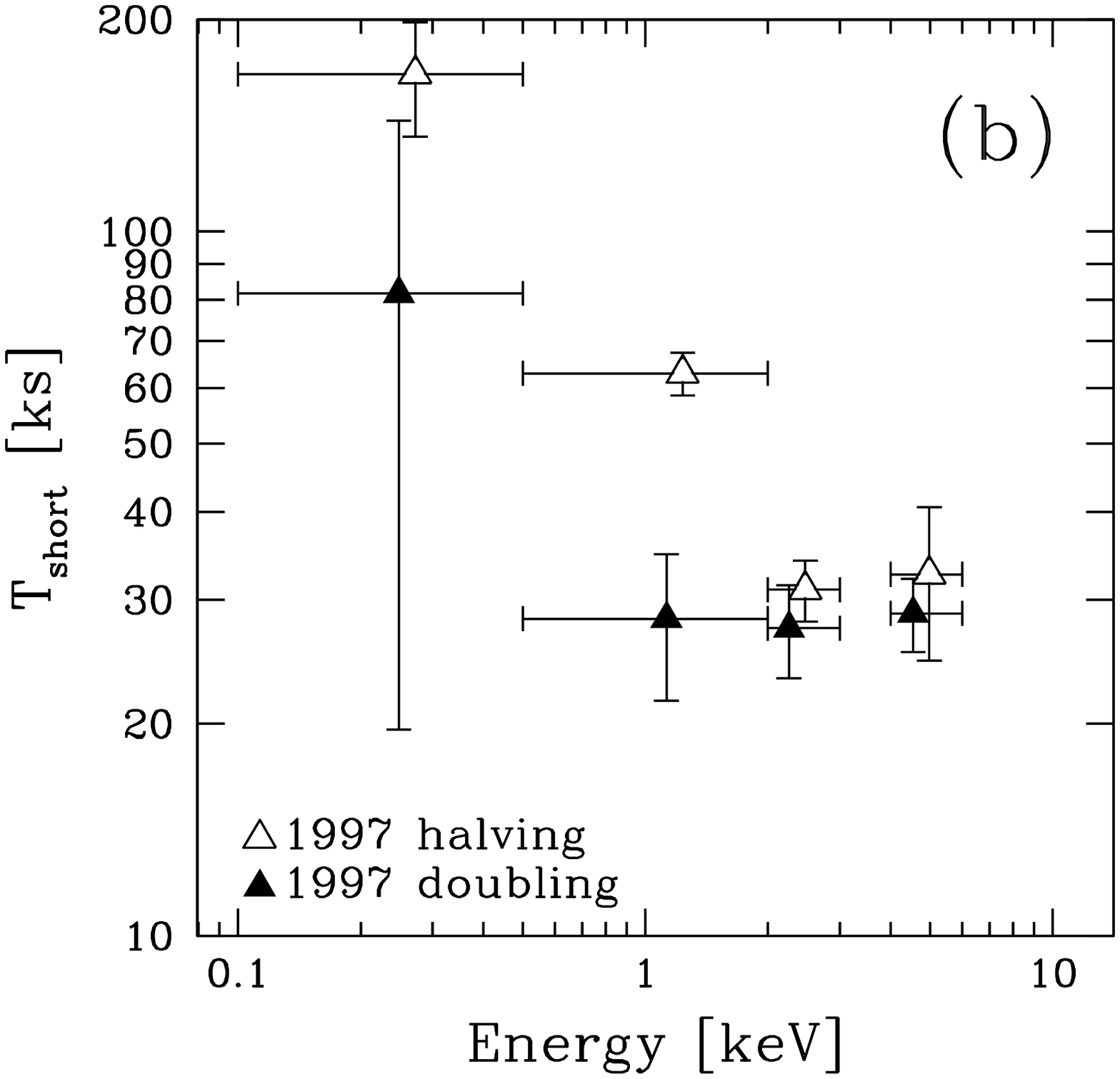}}
\vspace{-1.0cm}
\centerline{\includegraphics[width=0.75\linewidth]{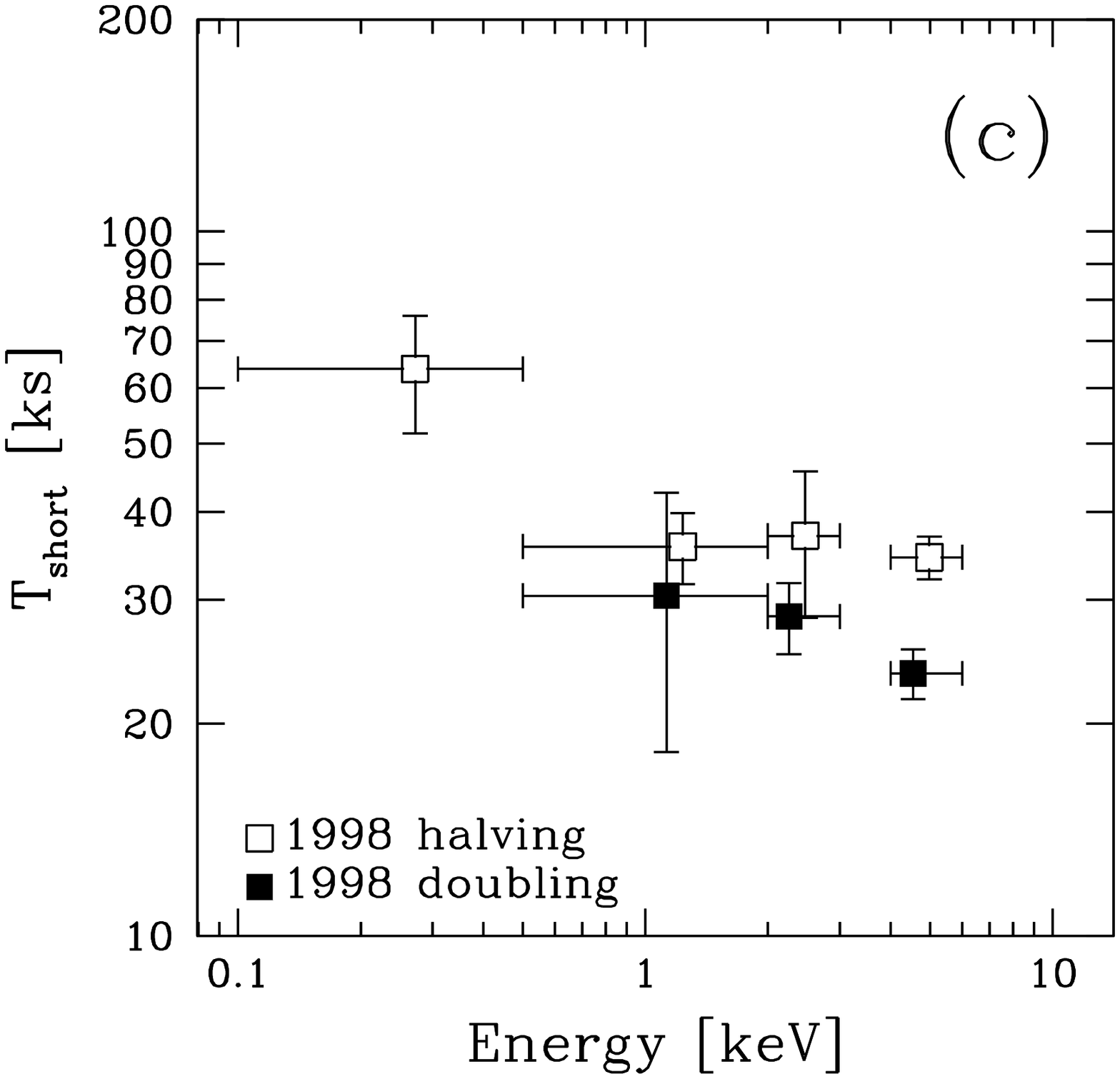}}
\vspace{-0.6cm}
\caption{\footnotesize\protect\baselineskip 8pt%
(a) comparison of global T$_\mathrm{short}$ for 1997 (crossed squares) and 1998
(diamonds), 
(b) comparison of doubling (black triangles) and halving (white triangles) timescales 
for 1997, and
(c) comparison of doubling (black squares) and halving (white squares) timescales 
for 1998 (see text).
}
\label{fig:t2_vs_energy}
\end{figure}

\subsubsection{Doubling/halving timescale}

Similarly, the \textit{``minimum halving/doubling''} timescale has
been computed for each energy band taking the average of values for
the different input light curve binning (but discarding the 200~s one
because too strongly subject to spurious results).  
As in Zhang et al. (\citealp{zhang_2155}) we rejected a
T$_{\mathrm{short},ij}$ if the fractional uncertainty was larger than 20~\%.

In order to minimize the contamination by isolated data
points we filtered the light curves excluding data lying at more than
3-$\sigma$ from the average computed over the 6 nearest neighbors (3 on
each side), and for the 500 s binning light curves we did not
consider pairs between data closer than 3 positions along the time
series.  Finally instead of taking the single absolute minimum value of
T$_{\mathrm{short},ij}$ (over all possible $i,j$ pairs)
T$_\mathrm{short}$ is determined as the average over the 5 lowest
values.  

The resulting T$_\mathrm{short}$ are listed at the bottom of
Table~\ref{tab:lc_fvar_t2} and shown in Figure~\ref{fig:t2_vs_energy}a.  
The main findings are: 
\begin{enumerate}
\item there is not significant difference between 1997 and 1998
observations, at any energy.
\item There is weak evidence that \textit{the softest X--ray band} exhibits
slower variability, while 
\item the timescales \textit{for the three higher} energy bands are
indistinguishable, and there is no sign of a trend with energy.
\end{enumerate}
We also distinguish  \textit{doubling} and  \textit{halving} timescales.
Unfortunately it has not been possible to obtain an estimate of the
\textit{doubling} time for the lowest energy band for 1998 data, because
there are only a few, uncertain, data during the rise of the flare.
We find that (see Table~\ref{tab:lc_fvar_t2}, and
Figure~\ref{fig:t2_vs_energy}b,c):
\begin{enumerate}
\setcounter{enumi}{3}
\item again the softest X--ray band shows longer (halving) timescales;
\item there is marginal evidence that the \textit{doubling} timescale is
shorter than the \textit{halving} one (i.e. the rise is faster than the fall).
\end{enumerate}
This latter result seems to hold both for 1997 and 1998 light curves 
and for each single
trial binning time, and for each energy (except in 5 cases out of 35)%
\footnote{We stress here that although the individual 
timescales are statistically 
consistent with each other, the behavior appear to be 
systematic at all energies.}.
However it is strongly weakened by the averaging over the several time binnings
because T$_\mathrm{short}$ slightly increases with the duration of the time
bins, thus yielding a larger uncertainty on the average.

\section{1998 detailed temporal analysis}
\label{sec:timing98}

We performed a more detailed analysis of the characteristics of the
April 21$^\mathrm{st}$ flare, in particular to determine any
significant different behavior  among different energy bands.  Three are the
main goals: i) to measure the flare exponential (or power--law) decay
timescales (or power--law index); ii) to evidence possible differences in
the flare rise and decay timescales; iii) to look for time lags among
the different energies.

We considered the energy ranges already adopted in
Section~\ref{sec:light_curves}.

\subsection{Decay Timescales}
\label{sec:timing98:decay_timescales}

A first estimate of the decay timescales has been presented by
Maraschi et al. (\citealp{maraschi_letter}), where a simple exponential
law has been fit to the decaying phase of the light curves, and a
dependence of the e--folding timescale on the energy band has been found.
Here we analyze the issue of determining the timescale in more detail and
in particular we consider the requirement for the presence of an underlying
steady emission.

In particular, the decay timescales for different energy bands have been
estimated by fitting the post--flare light curves with an exponential
decay superimposed to a constant (quasi--steady) flux
eventually constrained by
the fit (we immediately stress that an additional non--zero steady emission
is necessary in order to obtain a meaningful fit with an exponential decay). 
In one set of fits the underlying contribution is set to zero.  
More precisely, we performed the fits with three possible conditions for
the level of the steady contribution: un--constrained, constrained and zero.

The analytical expression of the function used in the fits is:
\begin{eqnarray*}
\mathrm{F}(t)&\;=\;&\mathrm{F}_\mathrm{steady} + \mathrm{F}_\mathrm{flaring}\; e^{-\frac{\mathrm{t}\; -\; \mathrm{T}_\mathrm{ref}}{\tau}} \\
             &\;=\;&\mathrm{F}_\mathrm{steady}\; \left(1 + {\cal R}\; e^{-\frac{\mathrm{t}\; -\; \mathrm{T}_\mathrm{ref}}{\tau}}\right)
\end{eqnarray*}
The model parameters are then the decay timescale $\tau$, the absolute
value of F$_\mathrm{steady}$, and the ratio ${\cal R}$ between the flaring
and the steady component taken at a reference time T$_\mathrm{ref}$, which
ideally should correspond to the peak of the flare.

The ``reference time'' T$_\mathrm{ref}$ for April 21$^\mathrm{st}$ flare
was T$_\mathrm{ref} = 25\times 10^3$~s (T$=0=10924.0$ TJD), and
T$_\mathrm{ref} = 170 \times 10^3$~s for April 23$^\mathrm{rd}$.
Note that the value of ${\cal R}$ obtained from
the fit to the second dataset is only a lower limit on the actual flare
amplitude since we do not have an estimate of the time at which the peak of
this putative second flare occurred.

\begin{figure}[!t]
\centerline{\includegraphics[width=0.80\linewidth]{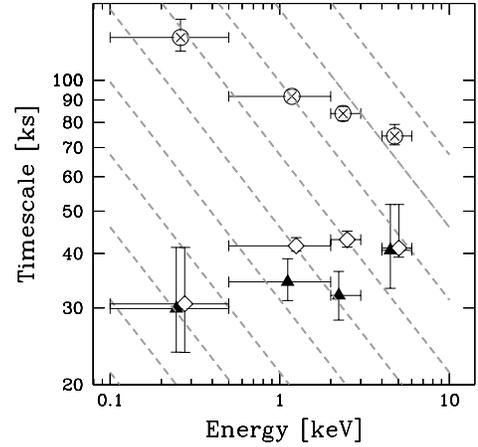}}
\vspace{-0.7cm}
\caption{\footnotesize\protect\baselineskip 8pt%
Decay timescales $\tau$ plotted versus the barycentric energy of the data
from which they have been derived.  The different symbols are: crossed
circles for zero--baseline, black triangles for the
un--constrained baseline, and diamonds for the constrained
baseline case.  The dashed lines show the loci of the relation $\tau
\sim \mathrm{E}^{-1/2}$, that holds if pure radiative cooling is
responsible for the spectral variability during the burst decay.  }
\label{fig:tau_vs_energy} 
\end{figure}

In any case, as our focus is actually on the properties of the first,
well defined, outburst, we used the parameters obtained from the April
23$^\mathrm{rd}$ data only to constrain the contribution from the
steady component. 
Furthermore, it turns out that the determination of the timescale $\tau$ is
not affected by the uncertainty on its contribution (see
Table~\ref{tab:lc_fits} and Fig.~\ref{fig:tau_vs_energy}).

\begin{figure}[t]
\centerline{\includegraphics[width=0.80\linewidth]{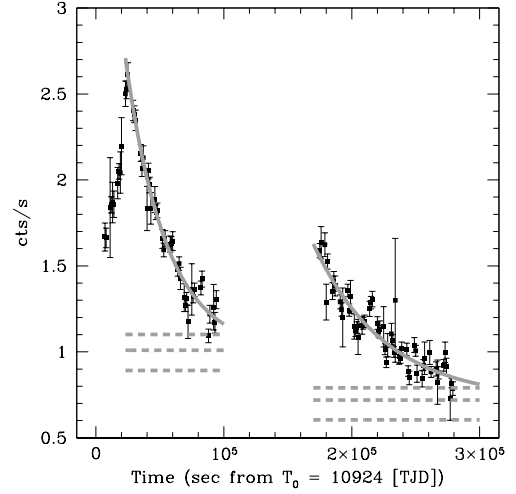}}
\vspace{-0.2cm}
\caption{\footnotesize\protect\baselineskip 8pt%
Example of the fit of the light curve with a constant+exponential decay.
The 2--3~keV data are shown, with a 1000 s time bin, together
with the best fit model (solid line through the data points), and the
best fit level of the steady component (three dashed lines corresponding
to the best fit value and the 90\% confidence interval, for two
interesting parameters).
}
\label{fig:lc98tau_example}
\end{figure}

We adopted a 1000~s binning for the 0.5--2~keV and 2--3~keV data, which
have better statistics, while we the binning time for the 0.1--0.5~keV and
4--6~keV light curves is of 2000 and 1500~s, respectively.
The analysis has been performed in two stages, for each energy band: 
\begin{itemize}
\item[i)] Fit to the April 21$^\mathrm{st}$ and 23$^\mathrm{rd}$ datasets
independently, yielding values of $\tau$, F$_\mathrm{steady}$ and ${\cal R}$. 
As an example, in Figure~\ref{fig:lc98tau_example} the 2--3~keV light curve
is shown together with its best fit models for April 21$^\mathrm{st}$ and
23$^\mathrm{rd}$.
\item[ii)] Fit of the April 21$^\mathrm{st}$ post--flare phase constraining
the level of the underlying steady component to be below the level of 
April 23$^\mathrm{rd}$ (the actual constraint is the upper bracket for a
two parameters 90\% confidence interval). 
We assume that the underlying emission varies on a timescale longer than
that spanned by the observations.
\end{itemize}
As anticipated we also modeled the decay of the April 21$^\mathrm{st}$ flare
setting to zero the offset.

The results are summarized in Tables~\ref{tab:lc_fits}
and~\ref{tab:decay_comparison}. 
The best fit $\tau$ for all the three cases are plotted in
Fig.~\ref{fig:tau_vs_energy} as a function of energy. 

The 0.1--0.5~keV timescale is not affected by the constraint on the
baseline flux and the 4--6~keV confidence interval only suffers a minor
cut (smaller errorbar), while both the 0.5--2 and 2--3~keV parameters are
significantly changed because the best fit un--constrained level of
the steady contribution is significantly higher than allowed by the April
23$^\mathrm{rd}$ data (see for instance Fig.~\ref{fig:lc98tau_example}).

\begin{deluxetable}{cccccc}
\tablecolumns{6}
\tablewidth{0.8\textwidth}
\tablecaption{Parameters of fit to light curves\tablenotemark{a,b}\label{tab:lc_fits}}
\tablehead{
\colhead{}&&
\colhead{LECS 0.1--0.5}&\colhead{LECS 0.5--2}&
\colhead{MECS 2--3}&\colhead{MECS 4--6}
}
\startdata
\multicolumn{6}{c}{constant + exponential decay} \\[0.1cm] \hline
\multicolumn{6}{l}{\underbar{April 21, zero--baseline}} \\[0.2cm]
$\tau$\tablenotemark{c} & (ks) &
125.3$^{+8.6}_{-5.8}$ & 91.8$^{+1.9}_{-1.5}$ &
83.8$^{+1.8}_{-2.2}$ & 74.4$^{+3.1}_{-2.3}$ \\[0.1cm]
  %
\multicolumn{6}{l}{\underbar{April 21, un-constrained}} \\[0.2cm]
${\cal R}$\tablenotemark{d} && 
\phn0.77$^{+0.24}_{-0.13}$ & \phn1.38$^{+0.19}_{-0.15}$ & 
\phn1.58$^{+0.22}_{-0.16}$ & \phn2.97$^{+3.01}_{-0.77}$ \\[0.1cm]
F$_\mathrm{steady}$ & (cts/s) & 
\phn0.68$^{+0.05}_{-0.08}$ & \phn2.23$^{+0.12}_{-0.18}$ & 
\phn1.01$^{+0.07}_{-0.09}$ & \phn0.39$^{+0.09}_{-0.12}$ \\[0.1cm]
$\tau$\tablenotemark{c} & (ks) &
29.9$^{+11.4}_{\phn-6.2}$\phn & 34.4$^{+4.5\phn}_{-3.2}$\phn &
32.1$^{+4.3\phn}_{-4.0}$\phn & 40.7$^{+11.1}_{\phn-7.4}$\phn \\[0.1cm]
  %
\multicolumn{6}{l}{\underbar{April 21, constrained}} \\[0.2cm]
${\cal R}$\tablenotemark{d} &&
\phn0.77$^{+0.24}_{-0.13}$ & \phn1.66$^{+0.11}_{-0.04}$ & 
\phn2.19$^{+0.16}_{-0.04}$ & \phn2.97$^{+3.01}_{-0.49}$ \\[0.1cm]
F\tablenotemark{e}$_\mathrm{steady}$ & (cts/s) & 
\phn0.67$^{    }_{-0.07}$ & \phn1.94$^{    }_{-0.07}$ & 
\phn0.79$^{    }_{-0.03}$ & \phn0.39$^{    }_{-0.12}$ \\[0.1cm]
&&
[$<$ 0.72]    & [$<$ 1.94] &
[$<$ 0.79]    & [$<$ 0.39] \\[0.2cm]
$\tau$\tablenotemark{c} & (ks) &
30.7$^{+10.6}_{\phn-6.9}$\phn & 41.7$^{+1.8\phn}_{-1.2}$\phn &
43.0$^{+2.0\phn}_{-1.7}$\phn & 41.2$^{+10.6}_{\phn-1.9}$\phn \\[0.1cm]
  %
\multicolumn{6}{l}{\underbar{April 23, un-constrained}} \\[0.2cm]
${\cal R}$\tablenotemark{d} && 
\phn0.38$^{+0.09}_{-0.07}$ & \phn0.95$^{+0.27}_{-0.13}$ &
\phn1.25$^{+0.24}_{-0.16}$ & \phn1.35$^{+0.17}_{-0.11}$ \\[0.1cm]
F$_\mathrm{steady}$ & (cts/s) & 
\phn0.70$^{+0.02}_{-0.04}$ & \phn1.77$^{+0.16}_{-0.24}$ & 
\phn0.72$^{+0.06}_{-0.08}$ & \phn0.36$^{+0.03}_{-0.03}$ \\[0.1cm]
$\tau$\tablenotemark{c} & (ks) &
29.0$^{+22.0}_{-10.3}$\phn & 66.6$^{+21.1}_{-13.3}$\phn & 
55.6$^{+14.0}_{\phn-9.1}$\phn & 39.2$^{+8.6\phn}_{-6.0}$\phn \\
  %
\cutinhead{constant + power law decay} 
${\cal R}$\tablenotemark{d} && 
$>$~1.715 & $>$~15.15 &
$>$~23.04 & $>$~39.37 \\[0.1cm]
F$_\mathrm{steady}$\tablenotemark{e} & (cts/s) & 
$<$ 0.600    & $<$ 0.619 &
$<$ 0.489    & $<$ 0.175 \\[0.1cm]
$\eta$\tablenotemark{f} &&
0.54$^{+0.37}_{-0.18}$ & 0.54$^{+0.05}_{-0.02}$ & 
0.62$^{+0.16}_{-0.02}$ & 0.70$^{+0.09}_{-0.03}$ \\
\enddata
\tablenotetext{a}{\footnotesize
In the \textit{un-constrained} case the flux level of the steady component
is left free to adjust during the fit.
In the \textit{constrained} case its flux level 
is required to be less or equal to the 
upper value of the 90 \% confidence interval
obtained by the fit of the
April 23$^\mathrm{rd}$ light curve in the corresponding energy band.
The values reported within square brackets represent
the imposed upper limits on F$_\mathrm{steady}$
(we do not quote the upper error value because in all four cases the upper
boundary of the confidence interval simply coincides with the imposed upper
limit).
Quoted errors are for a 1-$\sigma$ confidence interval for one or two 
interesting parameters (i.e. $\Delta\chi^2=1.0$ or $2.3$) for the case
\textit{without} and \textit{with} the baseline, respectively.
}
\tablenotetext{b}{\footnotesize 
We do not list the $\chi^2$ values because they are not really meaningful
estimates of the goodness of the fit. 
This happens because the errors on each bin of the light curves are very
small and there is significant variability on very short timescales which
of course is not accounted for by the simple exponential decay and
eventually it has the effect of giving an artificially high $\chi^2$ value.
For the fits of April 21$^\mathrm{st}$, the values of reduced $\chi^2$ are
between 1.6 and 4.1 for the cases including the baseline (for both
exponential and power law decays), and between 2.6 and 6.5 for the pure
exponential (zero--baseline) case.  The number of degrees of freedom (for
April 21$^\mathrm{st}$ datasets) is between 20 and 35, depending on the
energy band.}
\tablenotetext{c}{\footnotesize 
Timescale of the exponentially decaying flaring component.}
\tablenotetext{d}{\footnotesize 
Ratio between the flaring and steady components at
T$_\mathrm{ref}$ (see text). }
\tablenotetext{e}{\footnotesize 
90~\% confidence upper limit, for $\Delta\chi^2 = 4.61$, for the level of the
steady component.}
\tablenotetext{f}{\footnotesize 
power law decay spectral index.  }
\end{deluxetable}

\begin{deluxetable}{cccccc}
\tablecolumns{6}
\tablewidth{0pc}
\tablecaption{Timescales for Different Exponential Decay 
Fits\tablenotemark{a}\label{tab:decay_comparison}}
\tablehead{
\colhead{}&&
\colhead{LECS 0.1--0.5}&\colhead{LECS 0.5--2}&
\colhead{MECS 2--3}&\colhead{MECS 4--6}
}
\startdata
\multicolumn{6}{c}{April 21, zero--baseline} \\[0.1cm] \hline
T$_\mathrm{end} = 100 \times 10^3$ &&
125.3$^{+8.6}_{-5.8}$ & 91.8$^{+1.9}_{-1.5}$ &
83.8$^{+1.8}_{-2.2}$ & 74.4$^{+3.1}_{-2.3}$ \\[0.1cm]
 T$_\mathrm{end} = \phn90 \times 10^3$ &&
 120.3$^{+8.2}_{-5.7}$ & 89.0$^{+2.0}_{-1.5}$ &
 80.3$^{+2.4}_{-1.7}$ & 72.0$^{+3.7}_{-2.0}$ \\[0.1cm]
 T$_\mathrm{end} = \phn80 \times 10^3$ &&
 116.2$^{+9.0}_{-6.1}$ & 84.9$^{+2.0}_{-1.7}$ &
 73.8$^{+2.6}_{-1.8}$ & 68.5$^{+3.8}_{-2.4}$ \\[0.1cm]
 T$_\mathrm{end} = \phn70 \times 10^3$ &&
 106.2$^{+7.8}_{-5.9}$ & 80.7$^{+2.0}_{-1.9}$ &
 70.8$^{+2.5}_{-2.1}$ & 68.5$^{+3.4}_{-3.0}$ \\[0.1cm]
T$_\mathrm{end} = \phn60 \times 10^3$ &&
\phn95.3$^{+9.5}_{-6.6}$ & 76.0$^{+2.7}_{-2.1}$ &
70.0$^{+3.6}_{-2.8}$ & 70.3$^{+5.9}_{-4.1}$ \\[0.2cm]
                                              %
${\cal P}$\tablenotemark{b} && $4 \times 10^{-2}$ & $6 \times 10^{-8}$ & $1 \times 10^{-5}$ & \nodata \\ 
                                              %
                                              %
\cutinhead{April 21, un-constrained baseline}
T$_\mathrm{end} = 100 \times 10^3$ &&
\phn29.9$^{+11.4}_{\phn-6.2}$ & 34.4$^{+4.5}_{-3.2}$ &
32.1$^{+4.3}_{-4.0}$ & 40.7$^{+11.1}_{\phn-7.4}$ \\[0.1cm]
 T$_\mathrm{end} = \phn90 \times 10^3$ &&
 \phn33.8$^{+14.8}_{\phn-8.4}$ & 37.7$^{+5.5}_{-4.1}$ &
 34.0$^{+5.6}_{-4.7}$ & 46.3$^{+16.5}_{-10.4}$ \\[0.1cm]
 T$_\mathrm{end} = \phn80 \times 10^3$ &&
 \phn26.9$^{+10.9}_{\phn-7.1}$ & 34.5$^{+5.6}_{-4.8}$ &
 38.6$^{+10.4}_{\phn-7.5}$ & 53.6$^{+19.6}_{-16.9}$ \\[0.1cm]
 T$_\mathrm{end} = \phn70 \times 10^3$ &&
 \phn32.3$^{+24.0}_{-11.1}$ & 34.6$^{+9.7}_{-5.7}$ &
 45.6$^{+26.5}_{-12.5}$ & 61.7$^{+12.0}_{-26.3}$ \\[0.1cm]
T$_\mathrm{end} = \phn60 \times 10^3$ &&
\phn32.6$^{+70.4}_{-14.3}$ & 26.3$^{+8.0}_{-5.0}$ &
24.2$^{+11.2}_{\phn-5.9}$ & 18.9$^{+14.4}_{\phn-6.2}$ \\[0.2cm]
${\cal P}$\tablenotemark{b} && \nodata & \nodata & \nodata & \nodata \\ 
                                              %
\cutinhead{April 21, constrained baseline}
T$_\mathrm{end} = 100 \times 10^3$ &&
\phn30.7$^{+10.6}_{\phn-6.9}$ & 41.7$^{+1.8}_{-1.2}$ &
43.0$^{+2.0}_{-1.7}$ & 41.2$^{+10.6}_{\phn-1.9}$ \\[0.1cm]
 T$_\mathrm{end} = \phn90 \times 10^3$ &&
 \phn33.3$^{+15.1}_{\phn-8.0}$ & 41.2$^{+3.4}_{-1.3}$ &
 42.0$^{+3.0}_{-1.5}$ & 46.2$^{+16.6}_{\phn-7.2}$ \\[0.1cm]
 T$_\mathrm{end} = \phn80 \times 10^3$ &&
 \phn26.6$^{+11.3}_{\phn-3.4}$ & 40.8$^{+3.3}_{-1.2}$ &
 40.6$^{+8.9}_{-1.7}$ & 53.1$^{+19.4}_{-14.3}$ \\[0.1cm]
 T$_\mathrm{end} = \phn70 \times 10^3$ &&
 \phn31.9$^{+24.4}_{\phn-8.8}$ & 40.5$^{+6.1}_{-1.4}$ &
 \phantom{\footnotesize 3}45.5$^{+26.2}_{\phn-7.1}$ & 63.5$^{\phn+9.3}_{-24.4}$ \\[0.1cm]
T$_\mathrm{end} = \phn60 \times 10^3$ &&
\phn32.7$^{+69.3}_{\phn-9.4}$ & 40.2$^{+5.0}_{-1.8}$ &
41.5$^{+9.6}_{-2.4}$ & 44.3$^{+23.0}_{\phn-3.9}$ \\[0.2cm]
${\cal P}$\tablenotemark{b} && \nodata & \nodata & \nodata & \nodata \\ 
                                              %
                                              %
\enddata
\tablenotetext{a}{\footnotesize 
The decay of the April 21$^\mathrm{st}$ flare is modeled with an
exponential function over a non-variable flux contribution.
The listed numbers are the characteristic $\tau$'s (expressed in
ks) of the exponential decay, as a function of energy and T$_\mathrm{end}$.
Quoted errors are for a 1-$\sigma$ confidence interval for one or two 
interesting parameters for the case
\textit{without} and \textit{with} the baseline, respectively.
The timescales for the T$_\mathrm{end} = 100$~ks case are the same of
Table~\ref{tab:lc_fits}.
}
\tablenotetext{b}{\footnotesize 
Probability that the values of the timescale $\tau$ obtained for the
different choices of T$_\mathrm{end}$ are drawn from a unique parent
distribution ($\chi^2$ with respect to the weighted
average), for those cases where there is a significant
rejection of the ``null hypothesis''.
}
 %
 %
\end{deluxetable}


The main findings are:
\begin{itemize}
\item[1)] the decay timescales depend critically on the presence/ absence of
a contribution by a non variable component.
This is true not only for the value themselves (for the case
\textit{without} baseline are between a factor 2 and 4 longer) but also for
the relationship between the timescales and energy.  
In fact:
\begin{itemize}
\item[1a)] In the cases with baseline, the timescales range between
30 and 45$\times 10^3$ s, and \textit{do not} show a clear (if any)
relationship with the energy, rather suggesting an achromatic post--flare
evolution.
In fact, according to a $\chi^2$ test, the values of $\tau$ for the 4
energies are consistent with coming from the same distribution (the so
called ``null hypothesis'').
\item[1b)] On the contrary, in the case of pure exponential decay the
timescales follow a weak inverse relationship with the energy, 
$\tau \propto E^{-0.18 \pm 0.02}$ (both when considering $\tau$ or $E$ as
the dependent variable).
The $\chi^2$ probability in favor of the ``null hypothesis'' is in this
case negligible.
\end{itemize}
\item[2)] the amplitude of the variability (as measured by ${\cal R}$)
positively correlates with the energy, as reported in many cases for HBLs
(e.g. Sambruna et al. \citealp{sambruna_exosat}, Ulrich et al. \citealp{umu97}).
\end{itemize}
  
\subsubsection{Is there a steady component ?}
\label{sec:timing98:test_baseline}

Variability timescales and their energy dependence carry precious
information on the physics of the source: the puzzling dependence of
the results on the hypothesis of the presence of a steady component
requires further analysis.

Here we only discuss a simple test of the robustness of the results of
previous section, that however provides  useful hints. 
This consists in determining $\tau$ for different choices of the end time
(T$_\mathrm{end}$) of the interval over which the fit is performed.  
The underlying idea is that if the shape of the decay is close to the one
we are assuming (exponential with/without an offset) one can expect that
the model parameters do not change much as a function of T$_\mathrm{end}$.  
On the contrary, if the chosen model provides only a poor description of
the actual decay characteristics, the parameters will likely take different
values depending on T$_\mathrm{end}$, unless of course we postulate a
process whose timescale changes with time.

For each of the three cases (constrained, unconstrained, zero--baseline)
we tried 5 different choices of the end time, namely T$_\mathrm{end} =
60$, 70, 80, 90 and 100$\times 10^3$ seconds (from T$_0$), the last
one corresponding to the entire duration of the data coverage on April
21$^\mathrm{st}$.  

To evaluate the likelihood that the 5 values of $\tau$ obtained for these
choices of T$_\mathrm{end}$ are drawn from the same (normal) distribution,
we considered the probability ${\cal P}$ of the $\chi^2$ computed with
respect to their weighted%
\footnote{To each measure we assigned as $\sigma$ (and in turn as weight $\equiv
1/\sigma^2$), the mean between its $+$ and $-$ uncertainties.}
average%
\footnote{It should be noted that the weighted average has the
property of minimizing the $\chi^2$ for a dataset, and thus the
resulting values of probability that the data do belong to a unique
distribution are upper limits.}.

In Table~\ref{tab:decay_comparison} we summarize the results obtained
under each of the different hypotheses. 
There are a few points to note: 
\begin{enumerate}
\item in the \textit{zero--baseline} hypothesis the timescales obtained
changing T$_\mathrm{end}$ are \textit{not consistent} with being several
different realizations from a unique distribution.
Furthermore the clear trend of (decreasing) $\tau$ with (decreasing)
T$_\mathrm{end}$ supports the possibility that the variation of $\tau$
is significant.
\item On the contrary in both cases \textit{with} baseline the values of
$\tau$'s for the 0.1--0.5, 0.5--2 and 2--3~keV bands are very close
for all values of T$_\mathrm{end}$.
For the 4--6~keV band the scatter is somewhat larger, likely due to the
fact that at this energy there is the largest amplitude of variability,
i.e. the smallest contribution of the putative baseline.
The same kind of effect of lowered sensitivity to the details of a putative
baseline, is responsible for the good result of the 4--6~keV band in the
case \textit{without} baseline.
\item Even in the case \textit{without} baseline the values of $\tau$ for
the various energy bands become very similar for T$_\mathrm{end} = 60
\times 10^3$~s.
Therefore even if the failure of the test by the pure exponential
decay model could be interpreted as the signature of a mechanism whose
characteristic timescale changes with time (unlike a true exponential decay),
there is anyway an indication that in the earlier stages of the decay the
evolution is not energy dependent.
\end{enumerate}
On the basis of this simple consistency check we can therefore infer that
\textit{if the decay were exponential} there is evidence for the presence
of an underlying non--variable component. 
Thus the post--flare phase does not show any dependence on energy, and the
timescale is of the order of 30--40~ks.

\begin{figure}[t]
\centerline{\includegraphics[width=0.80\linewidth]{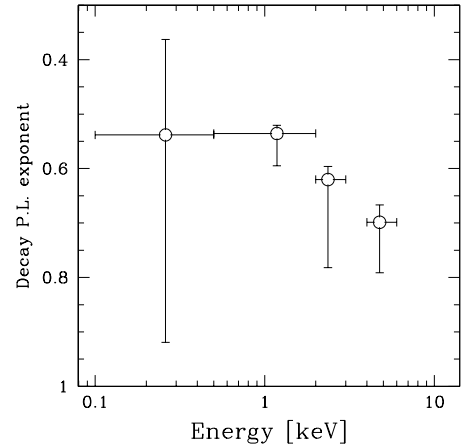}}
\vspace{-0.7cm}
\caption{\footnotesize\protect\baselineskip 8pt%
Index of the power law decay plotted versus energy.
Only the value for the April 21 post peak decay is plotted.
}
\label{fig:eta_vs_energy}
\end{figure}

\subsection{Power Law Decay}
\label{sec:timing98:power_law}

As an alternative description of the decay of April 21$^\mathrm{st}$ flare
we consider a power law, where F$_\mathrm{flaring} \propto 
(\mathrm{T}/\mathrm{T}_\mathrm{ref})^{-\eta}$,
again leaving the possibility of an offset from a steady component.
The results are summarized in Table~\ref{tab:lc_fits} and
Figure~\ref{fig:eta_vs_energy}.

A fundamental difference with respect to the exponential decay case is
that an additional steady component is never required: the power
law fits yield only (not very stringent) upper limits on
F$_\mathrm{steady}$ and only for the 0.1--0.5~keV band the formal best fit
value for F$_\mathrm{steady}$ is different from zero
(F$_\mathrm{steady}= 0.26$ cts/s). 

Again there is only a weak relationship of the decay properties with
energy, with a marginal indication that the flux decrease is faster at
higher energies.
The values for the 4 energy bands are consistent with a common (weighted
average) value of $\eta = 0.583$, while for the subset comprising only the
three higher energy datasets the probability for the ``null hypothesis''
is reduced to about 0.07.

\subsection{Rise vs. Fall}
\label{sec:timing98:rise_vs_fall}

A further interesting question is whether the rise and the decay phases of
the flare are characterized by the same timescale (i.e. if the flare is
symmetric or asymmetric) and in particular how these properties might
correlate with energy.

The observational coverage of the rise is not good enough to apply the
same direct technique used for the decay
(\S\ref{sec:timing98:decay_timescales}), to constrain the timescale
of an exponential rise, with all the parameters left free. Thus the
flux level of the steady component is fixed at the best fit value
obtained for the decay, or set to 0.

\begin{deluxetable}{lcccccc}
\tablecolumns{7}
\tablewidth{0pc}
\tablecaption{Difference between Rise and Fall Timescales\tablenotemark{a}\label{tab:rise_vs_fall}}
\tablehead{
\colhead{case\tablenotemark{b}}&&
\colhead{LECS 0.1--0.5}&\colhead{LECS 0.5--2}&
\colhead{MECS 2--3}&\colhead{MECS 4--6}&\colhead{${\cal P}$\tablenotemark{c}}
}
\startdata
 %
 %
F$_\mathrm{steady} = 0$ &&
2.9$^{+33.4}_{-12.3}$ & 26.3$^{+15.0}_{\phn-3.4}$ &
33.1$^{+11.2}_{\phn-2.5}$ & 38.6$^{+4.8}_{-3.4}$ &
\nodata \\[0.2cm]
F$_\mathrm{steady}$ best fit un-constrained &&
1.2$^{+6.9\phn}_{-4.3}$ & 13.7$^{+4.3\phn}_{-1.9}$ &
14.8$^{+7.9\phn}_{-1.3}$ & 23.6$^{+8.5}_{-1.9}$ & 
$3 \times 10^{-2}$ \\[0.2cm]
F$_\mathrm{steady}$ best fit constrained &&
1.3$^{+6.2\phn}_{-4.3}$ & 16.1$^{+6.3\phn}_{-2.1}$ &
19.8$^{+10.9}_{\phn-1.6}$ & 24.6$^{+9.1}_{-2.2}$ & 
$1 \times 10^{-2}$ \\[0.1cm]
 %
 %
\enddata
\tablenotetext{a}{\footnotesize 
The values of $\Delta \mathrm{T} = \mathrm{T}_\mathrm{fall} -
\mathrm{T}_\mathrm{rise}$ have been computed for two different binning of
the light curves, 500 and 1000 s, and the average is shown (in ks).
Confidence intervals are 1--$\sigma$ for 1 interesting parameter, 
because the baseline was not free to vary.
The upper errorbars (giving the largest possible value of $\Delta \mathrm{T}$) 
derive from the upper boundary of the T$_\mathrm{fall}$ confidence interval
and the lower boundary of the T$_\mathrm{rise}$ confidence interval.
The lower errorbars (giving the lowest possible value of $\Delta \mathrm{T}$) 
derive from the lower boundary of the T$_\mathrm{fall}$ confidence interval
and the upper boundary of the T$_\mathrm{rise}$ confidence interval.
}
\tablenotetext{b}{\footnotesize 
In each case the level of the steady component was kept fixed at the best
fit value for each energy band, as listed in Table~\ref{tab:lc_fits}.
}
\tablenotetext{c}{\footnotesize 
Probability that the four values of $\Delta \mathrm{T}$ are drawn from the
same parent distribution, computed using the $\chi^2$ of the data with
respect to their weighted average.
}
\end{deluxetable}


\begin{figure}[t]
\centerline{\includegraphics[width=0.80\linewidth]{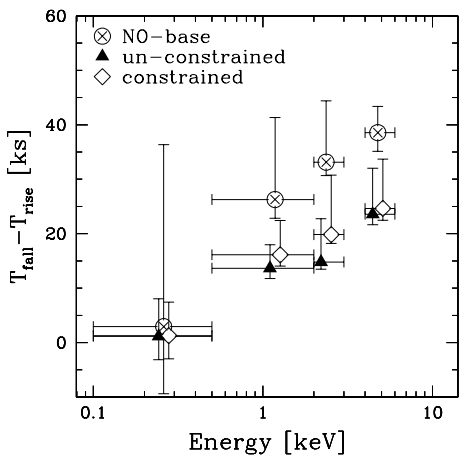}}
\vspace{-0.4cm}
\caption{\footnotesize\protect\baselineskip 8pt%
Difference between the post--peak decay and the pre--peak rise timescales,
plotted versus energy.
The different symbols are: crossed circles for F$_\mathrm{steady} = 0$ case,
black triangles for the F$_\mathrm{steady}$ set at the best fit value of
the un--constrained case, and diamonds for F$_\mathrm{steady}$ set 
at the best fit value obtained for the April 23$^\mathrm{rd}$ data (see
Table~\ref{tab:lc_fits}).
}
\label{fig:asymmetry}
\end{figure}

We are mainly interested in the comparison of the properties of the
outburst just before and after the top of the flare (see discussion in
Paper~2), and the interval of about 17--18 ks preceding the
flare is long enough for this purpose.

We thus focused on the ``tip'' of the flare, considering data on the decay
side only up to the time when the flux reaches approximately the level at
which the observation starts.
The duration of the post--peak intervals needed to decrease the flux to the
pre--flare level are of about 15--20, 25, 30--35 and 40~ks for the four
energy bands.
Since the uncertainties on the decay time of the light curves are not
larger than 5~ks, the differences in duration and the trend (longer time at
higher energy) are real.

The differences between rise and fall timescales, estimated with fits to
the light curves, are shown in Table~\ref{tab:rise_vs_fall}, and plotted in
Figure~\ref{fig:asymmetry}.  
The flare is symmetric for the softest energy band, while it is definitely
asymmetric in the other three bands, with an asymmetry which appears to
increase systematically with energy: at higher energies the rise phase is
increasingly faster than the decay (the significance is reported in
Table~\ref{tab:rise_vs_fall}).  

It is worth stressing that these numbers indicate only relative changes 
in the timescales before and after the peak of the flare.
On the other hand, since we found that the decay timescales are very similar
once the baseline flux is taken into account
(\S\ref{sec:timing98:decay_timescales}), this asymmetry probably means
that the rise times get shorter with increasing energy.

\subsection{Time Lag}
\label{sec:timing98:lag}

We performed a detailed cross correlation analysis using two different
techniques suited to unevenly sampled time series:
the Discrete Correlation Function (DCF, Edelson \& Krolik 
\citealp{edelson_krolik88}) and the Modified Mean Deviation (MMD, Hufnagel
\& Bregman \citealp{hufnagel_bregman92}).
Moreover, Monte Carlo simulations, taking into account ''flux
randomization'' (FR) and ''random subset selection'' (RSS) of the data
series (see Peterson et al. \citealp{peterson_etal98}), were used to
statistically determine the significance of the time lags between different
X--ray bands obtained with the DCF and MMD.
We refer for the relevant details of such analysis to
Zhang et al. (\citealp{zhang_2155}). 

\begin{figure}[t]
\centerline{\includegraphics[width=0.65\linewidth,angle=-90]{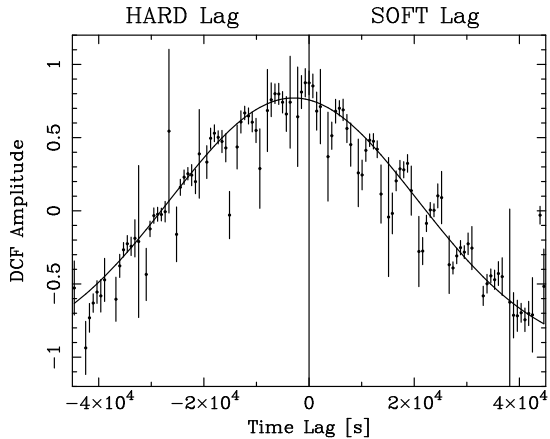}}
\caption{\footnotesize\protect\baselineskip 8pt%
Result of the cross--correlation DCF method applied to the 0.1--1.5 and
3.5--10~keV light curves.
Negative values corresponds to the 3.5--10~keV band lagging the 0.1--1.5~keV one.
The continuous curve superimposed to the data is the best fit with a 
constant$+$Gaussian model.
The actual fit has been performed only to the ``core'' of the data, i.e.
taking into account the data in the interval ranging from $-$30000
to 30000 seconds.
}
\label{fig:dcf_mmd}
\end{figure}

\begin{figure}[t]
\centerline{\includegraphics[width=0.80\linewidth]{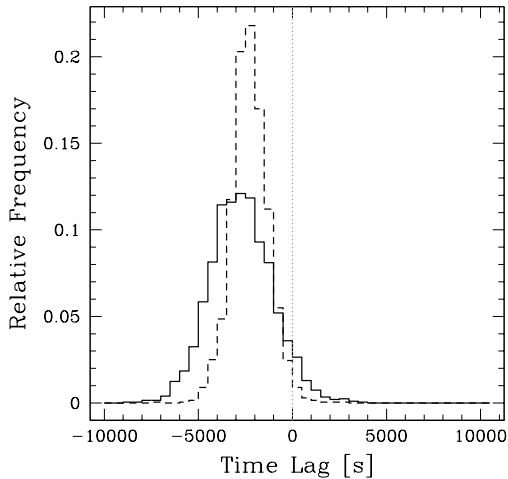}}
\vspace{-0.4cm}
\caption{\footnotesize\protect\baselineskip 8pt%
Cross Correlation Peak Distributions for the FR/RSS simulations.
The solid line is for the analysis with DCF technique, and the dashed for the
case of MMD.
Negative values corresponds to the 3.5--10~keV band lagging the 0.1--1.5~keV one.
}
\label{fig:ccpd}
\end{figure}

We binned the light curves over 300~s in the 0.1--1.5 and 3.5--10 keV
bands (whose effective barycentric energies are E $\simeq 0.8$ and E $\simeq
5$~keV), and a trial time step of 720 s is adopted for both the DCF
and MMD.
DCF amplitude versus time lag is plotted in Figure~\ref{fig:dcf_mmd}.
A negative value means that variations in the 3.5--10~keV band light curve
lag those occurring in the  0.1--1.5~keV one (i.e. hard lag).
The best Gaussian fits for both DCF and MMD result in negative time
lags of $-2.8 \pm 0.2$ (DCF) and $-2.1 \pm 0.3$ (MMD)~ks, indicating
that the medium energy X--ray photons lag the soft ones.

The Cross Correlation Peak Distribution (CCPD) obtained from the
FR/RSS Monte Carlo simulations is shown in Figure~\ref{fig:ccpd} both
for the DCF and the MMD methods.  The average lags resulting from CCPD
are $-2.7^{+1.9}_{-1.2}$ ks for DCF, and $-2.3^{+1.2}_{-0.7}$ ks for
MMD (1 $\sigma$ confidence intervals, two sided with respect to the
average), confirming the significance of the above results with high
($>$ 90~\%) confidence.

The total integral probabilities for a negative lag (that would be the
actual measure of the confidence of the ``discovery'' of the hard lag) 
are $\simeq 95.0$\% (DCF) and $\simeq 98.7$\% (MMD).

\section{Summary of Variability Properties}
\label{sec:disc:variability}

We presented a comprehensive temporal analysis of the flux variability
characteristics in several energy bands, of \textit{Beppo}SAX
observations of Mkn~421.  
The primary results of this study are the following: 
\begin{enumerate}
\item The fractional \textit{r.m.s.} variability is higher at higher energies.
\item The fractional \textit{r.m.s.} variability does not change with the
brightness state of the source.
\item The \textit{minimum halving/doubling time} is
longer for the softest energy band.
\item The \textit{minimum halving/doubling time} at a given energy does not
change with the brightness state of the source.
\item There is a hint that the \textit{doubling} timescale is shorter than
the \textit{halving} timescale.
\end{enumerate}
The findings 1 and 2, on F$_\mathrm{var}$, confirm the well known
(although not fully explained) phenomenology of HBLs (see Ulrich et al.
\citealp{umu97} and references therein).

The results on the variability timescales (findings 3, 4, 5) show good
agreement with those found with the more thorough analysis of the 1998
flare, that can be summarized as:
\begin{enumerate}
\setcounter{enumi}{5}
\item The flare decay is consistent with being achromatic, both if modeled
as an exponential decay with an additional contribution from a steady
component and in the case of a power law decay.
\item The results on timescales are at odds with the simplest
possibility of interpreting the decay phase, i.e. that it is driven by
the radiative cooling of the emitting electrons.  This would give
rise, in the simplest case, to a dependence of the timescale on
energy, $\tau \sim \mathrm{E}^{-1/2}$.  The tracks
corresponding to this relationship are overlaid to the data in
Fig.~\ref{fig:tau_vs_energy}, and it is clear that they can not be
reconciled.  
\item The harder X--ray photons lag the soft ones, with a
delay of the order of few ks.
This finding is \textit{opposite} to what is normally found in the best
monitored HBL where the soft X--rays lag the hard ones 
(e.g.  Takahashi et al. \citealp{takahashi96} for Mkn~421; 
Urry et al.  \citealp{urry_2155}, 
Zhang et al. \citealp{zhang_2155}, and 
Kataoka et al.  \citealp{kataoka_2155} for PKS 2155--304;
Kohmura et al. \citealp{kohmura} for H0323$+$022).
The latter behavior is usually interpreted in terms of cooling of
the synchrotron emitting particles. 
\item Possible asymmetry in the rise/decay phases: the flare seems to be
symmetric at the energies corresponding (roughly) to the peak of the 
synchrotron component (see the spectral analysis in Paper II),
while it might have a faster rise at higher energies.
This could be connected to the observed hard--lag.
\item Finally, the characteristics of the April 21$^\mathrm{st}$ flare 
suggest the presence of a quasi--stationary emission contribution, which
seems to be dominated by a highly variable peaked spectrum.
\end{enumerate}


\section{Conclusions}
\label{sec:conclusions}

\textit{Beppo}SAX has observed Mkn~421 in 1997 and 1998.  
We analyzed and interpreted the combined spectral and temporal evolution in
the X--ray range.  
During these observations the source has shown a large variety of
behaviors, providing us a great wealth of information, but at the same time
revealing a richer than expected phenomenology.

In this paper we have presented the first part of the analysis, focused on the
study of the variability properties.

The fact that the higher energy band lags the softer one (with a delay of
the order of 2--3~ks) and the energy dependence of the shape of the light
curve during the flare (with faster flare rise time at higher energies) provide
strong constraints on any possible time dependent particle acceleration
mechanism.
In particular, if we are indeed observing the first direct signature of
the ongoing particle acceleration, progressively ``pumping''
electrons from lower to higher energies, the measure of the delay between
the peaks of the light curves at the corresponding emitted frequencies would 
provide a tight constraint on the timescale of the acceleration process.

The decomposition of the observed spectrum into two components 
(a quasi stationary one and a peaked, highly variable one) might allow 
us to determine the nature and modality of the
energy dissipation in relativistic jets.

In Paper~II we complement these findings with those of the time
resolved spectral analysis and develop a scenario to interpret the complex
spectral and temporal phenomenology.

\acknowledgments

We are grateful to the \textit{Beppo}SAX Science Data Center (SDC) for their 
invaluable work and for providing standardized product data archive, 
and to the \textit{Rossi}XTE ASM Team.
We thank Gianpiero Tagliaferri and Paola Grandi for their contribution to
our successful \textit{Beppo}SAX program, and for useful comments
and the anonymous referee for useful suggestions which have 
improved the clarity of the paper. 
AC, MC and YHZ acknowledge the Italian MURST for financial support. 
This research was supported in part by the National Science Foundation
under Grant No.~PHY94--07194 (AC).
Finally, GF thanks Cecilia Clementi for providing tireless stimulus.


\appendix
\section{A. Definitions of F$_{var}$ and T$_\mathrm{short}$}
\label{sec:appendix:estimators}

The fractional $r.m.s.$ variability amplitude is a useful parameter to
characterize the variability in unevenly sampled light curves.
It is defined as the square root of the so called excess variance 
(e.g., Nandra et al. \citealp{nandra_97}).
This parameter, also known as the true variance (Done et al.
\citealp{done_92}), is computed by taking the difference between the
variance of the overall light curve and the variance due to measurement
error, normalized by dividing by the average squared flux (count rate). 

We consider a dataset $F_i(t_i)$ ($i=1,N$), with an uncertainty $\sigma_i$
assigned to each point.

The fractional \textit{r.m.s.} variability parameter is then defined as:
\begin{equation}
\mathrm{F}_\mathrm{var}  \;\equiv\; \frac{\left(\sigma^2_F - \Delta^2_F\right)}{\langle F\rangle}^{1/2},
\end{equation}
where 
\begin{eqnarray}
\sigma^2_F & \;\equiv\; & \frac{1}{N-1}\;\sum_{i=1}^{N} (F_i - \langle F\rangle)^2 \\
\Delta^2_F & \;\equiv\; & \frac{1}{N}\; \sum_{i=1}^{N} \sigma_i^2.
\end{eqnarray}

The definition of the \textit{``shortest variability timescale''} is the following:
\begin{eqnarray}
\mathrm{T}_\mathrm{short} &\;\equiv\;& \min_{\forall i,j}(\mathrm{T}_{\mathrm{short},ij}) \\ 
\mathrm{T}_{\mathrm{short},ij} &\;\equiv\;& \left| \frac{F_{ij}\;\Delta T_{ij}}{\Delta F_{ij}}\right|,
\end{eqnarray}
where $F_{ij} \;\equiv\; (F_i + F_j)/2$, $\Delta F_{ij} \;\equiv\; F_i - F_j$,
and $\Delta T_{ij} \;\equiv\; T_i - T_j$.



\end{document}